%% file: MoscaCorregido.tex
\documentclass[a4paper,twocolumn,12pt]{article}
\usepackage[spanish]{babel}
\usepackage{ae}
\usepackage[latin1]{inputenc} 
\usepackage[T1]{fontenc} 
\usepackage{layout}
\usepackage{array}
\usepackage{graphicx}
\usepackage{amsmath}
\usepackage{multicol}
\usepackage{rotating}  
\hyphenrules{nohyphenation} 

\usepackage[]{cite}   
\usepackage{anysize}  
\usepackage{endnotes} 

\usepackage{tikz,fp,ifthen,fullpage}
\usetikzlibrary{arrows,snakes,backgrounds}
\usepackage{pgfplots}
\usetikzlibrary{shapes,trees}
\usetikzlibrary{calc,through,backgrounds,decorations}
\usetikzlibrary{decorations.shapes,decorations.text,decorations.pathmorphing,backgrounds,fit,calc,through,decorations.fractals}
\usetikzlibrary{fadings,intersections}
\usetikzlibrary{patterns}
\usetikzlibrary{mindmap}
\marginsize{1.5cm}{1.0cm}{1.5cm}{2cm}

\begin{document}

\renewcommand{\tablename}{Tabla}
\renewcommand{\abstractname}{}
\renewcommand{\thefootnote}{\arabic{footnote}}

\title{
        {\LARGE Experimentos virtuales sobre una mosca vagabunda: más allá de la solución de Neumann}\\
        \footnotesize   \textit{(Virtual experiments about a walker fly: beyond the Neumann`s  solution)}
}
\author
{ 
Paco Talero$^{1,2}$,César Mora$^{2}$,Orlando Organista$^{1,2}$ y Fabian Galindo$^{3}$\\
$^{1}$ {\small Grupo F\'{\i}sica y Matem\'{a}tica, Dpt de Ciencias
Naturales, Universidad Central},\\ 
\small {Carrera 5 No 21-38, Bogot\'{a}, D.C. Colombia.},\\ 
$^{2}$ {\small Centro de Investigaci\'{o}n en Ciencia Aplicada y Tecnolog\'{\i}a Avanzada del Instituto Polit\'{e}cnico Nacional,} \\
\small {Av. Legaria 694, Col. Irrigaci\'{o}n, C. P. 11500, M\'{e}xico D. F.}\\
$^{3}$ {\small IED Jorge Soto del Corral,  calle 3 No 2-64 este SEDE Bogotá}
}
\date{}

\twocolumn
[
\begin{@twocolumnfalse}
\maketitle 
\begin{abstract}
Se extiende el problema tradicional que pregunta por la distancia que recorre una mosca con rapidez  constante  cuando se mueve entre dos trenes en movimiento uniforme rectilíneo con dirección de colisión, se revisa tanto la solución trivial como la atribuida a Neumann y mediante una simulación se amplía  el problema  para incluir trenes ace\-le\-ra\-dos. Este  problema  puede usarse  en cursos de física e\-le\-men\-tal para ilustrar la relación física-matemática, dar contexto al  análisis gráfico y estudiar diversas situaciones en cinemática unidimensional. Además, se resalta la fortaleza  didáctica del problema a través de la analogía que guarda con el funcionamiento de un sensor de velocidad basado en  pulsos de ultrasonido.\\
\textbf{Palavras-chave:} problema, mosca, Neumann, sensor, ultrasonido.\\ 

We extend the traditional problem of calculating the traveled distance of a fly that is moving with constant speed on a straight path between two trains that are approaching one another, we found two analytical solutions for the traditional problem and we propose a generalization to the accelerated motion of the trains through a simulation.  With this problem the teachers can teach some topics of basic physics as the relations between mathematical and physics, the graphical analysis and various situations on kinematics in one dimension. We show an analogy between the move of the fly and a ultrasonic wave pulse in a sensor of velocity, this show the pedagogical power of this problem. 

\end{abstract}

\end{@twocolumnfalse}
]

\section{Introducción}

Dada la complejidad de muchos  sistemas  en diversos campos de la física actualmente es frecuente que su estudio se lleve  a cabo mediante la  interacción entre teoría, experimento y simulación, intercambiando así ideas  que contribuyen a la comprensión física de tales sistemas.  De esta manera la simulación ha demostrado que las leyes  que des\-cri\-ben el  comportamiento de múltiples  sistemas  físicos pueden reducirse a algoritmos que al ser implementados en un ordenador determinan el comportamiento, por lo general aproximado, de estos  sistemas.  En particular,  la ejecución de una simulación  que obedece a tales algoritmos viene a ser como la realización de un experimento en un espacio virtual, donde es posible restablecer   parámetros,  ajustar condiciones iniciales, cambiar condiciones de frontera  y en general plantear  y confrontar diversas hipótesis de trabajo, por tales razones  suele llamarse a este tipo de procesos de investigación experimentos virtuales (ExV)\cite{Landau,Tobo,Tobo2,Pang,Esque,EXV,Wol}. No obstante, debe tenerse claro que los ExV son una manera de hacer física teórica y que el experimento real será siempre necesario.  

Hace unas cinco décadas los problemas concernientes a la física computacional correspondían fundamentalmente a cursos de posgrado, sin embargo durante los últimos $20$ años en carreras de ciencias e ingeniería ha habido un incremento considerable en el interés por plantear el estudio de esta rama de la física y asignaturas afines desde los primeros espacios de formación.   Así es como en  la física educativa los ExV han mostrado múltiples ventajas frente a los métodos tradicionales, pues concretizan algunos aspectos referentes a problemas abstractos; sirven como tutores en algunos procesos cognitivos;  permiten resolver problemas que no tienen una solución analítica; desarrollan en el estudiante  intuición física y lo  acercan  a la solución de problemas menos ideales y más cercanos a la realidad; brindan escenarios aproximados a la física computacional  a través de la solución de problemas complejos  con pocos conocimientos matemáticos y en general ofrecen un  contexto alternativo para la discusión de numerosos  conceptos físicos. Además, conviene resaltar que los ExV se implementan  en  física educativa fundamentalmente mediante lenguajes de programación, visualizadores gráficos, hojas de cálculo, calculadores simbólicos y software de animación\cite{ArcoI,Wilson,Timberlake,HojaC,Curri,Bernulli,Santa}.   

En este trabajo se extiende el problema tradicional de una mosca que se mueve con rapidez  cons\-tan\-te manteniendo una trayectoria  recta  entre dos trenes que se aproximan uno al otro con rapidez cons\-tan\-te\cite{Martin}, se muestran  dos  soluciones analíticas  y  mediante ExV se amplía el análisis a trenes ace\-le\-ra\-dos, también se usa el problema para explicar algunos  principios físicos asociados al funcionamiento de un sensor de velocidad. En la sección ($2$) se muestra una solución analítica e\-le\-men\-tal del problema tradicional, en la sección ($3$) se expone una solución tipo Neumann  que implica el uso de series de potencias, en la sección ($4$) se dota a los trenes con movimiento arbitrario y se resuelve el problema mediante un ExV basado en el  algoritmo de Euler  para el movimiento de la mosca, en la sección ($5$) se hace concreto el problema al explicar el funcionamiento de los sensores de velocidad basados en pulsos de ultrasonido y finalmente en la sección ($6$) se presentan las conclusiones.

\section{Solución e\-le\-men\-tal}

En el ejercicio  tradicional  planteado se con\-si\-de\-ran  dos trenes $A$ y $B$ que se acercan uno al otro con rapidez cons\-tan\-te $v_A$ y $v_B$  en una vía recta, entre ellos una mosca se mueve también con rapidez cons\-tan\-te $v_m$  mayor que la de los trenes, la mosca toca un tren y retorna al otro  sin pérdida de tiempo ni variación apreciable en su rapidez durante el cambio de dirección, la mosca continúa oscilando entre los trenes hasta que estos colisionan. Generalmente la actividad propuesta consiste en calcular la distancia recorrida por la mosca \cite{Martin}.
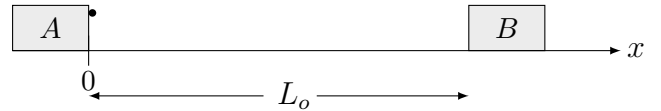
\begin{figure}[ht]
\begin{center}
\begin{tikzpicture}[xscale=1.0,yscale=1.0]
\fill[gray!15](0,0) rectangle (1,0.6);
\draw[color=black] (0,0) rectangle (1,0.6);
\node[] at (0.5,0.3) {\small{$A$}};
\fill[gray!15] (6,0) rectangle (7,0.6);
\draw[color=black] (6,0) rectangle (7,0.6);
\node[] at (6.5,0.3) {\small{$B$}};
\draw[-latex, black] (0cm,0cm)--(8cm,0cm);
\draw[black] (1cm,-0.2cm)--(1cm,0.2cm);
\node[] at (1.0,-0.4) {\small{$0$}};
\node[] at (8.2,0.0) {$x$};
\fill[black] (1.05cm,0.5cm) circle(0.05cm); 
\draw[-latex, black] (3.3cm,-0.6cm)--(1.0cm,-0.6cm);
\draw[-latex, black] (4.0cm,-0.6cm)--(6.0cm,-0.6cm);
\node[] at (3.7,-0.6) {$L_{o}$};
\end{tikzpicture}
\caption{Configuración inicial del sistema.}
\label{moscaA}\end{center}
\end{figure}

El sistema se simplifica  al considerar los trenes como rectángulos ideales y la mosca como un punto matemático. De esta manera, en la Fig.\ref{moscaA} se ilustra la configuración inicial del sistema, aquí  los trenes se encuentran entre sí a una distancia $L_o$ y la mosca parte del tren $A$ al $B$. De acuerdo con lo anterior  las ecuaciones de posición contra tiempo de los trenes y la mosca son
\begin{equation}\label{posTA}
 x_{A} =v_{A}t,
\end{equation}
\begin{equation}\label{posTB}
x_{B}=L_{o}-v_{B}t,
\end{equation}
y
\begin{equation}\label{posM}
x_{m}=v_{m}t.
\end{equation}
Donde $v_A$, $v_B$ y $v_m$ son positivas ya que representan el modulo de cada velocidad. Ahora, Para encontrar la distancia que recorre la mosca desde que parte del tren $A$ hasta que los trenes colisionan es necesario calcular el tiempo de colisión  $t_{c}$ entre los trenes. De las ecuaciones (\ref{posTA}) y (\ref{posTB}) se obtiene
\begin{equation}\label{tiem}
t_{c}=\frac{{L_{0}}}{v_{A}+v_{B}}.
\end{equation}
Ahora se usa este tiempo para calcular la distancia que la mosca se movió, al usar las ecuaciones  (\ref{posM}) y (\ref{tiem}) se encuentra
\begin{equation}\label{dis}
D=\frac{{v_{m}L_{0}}}{v_{A}+v_{B}},
\end{equation}
que es la distancia buscada.

\section{Solución tipo Neumann}
Se dice popularmente, quizá simplemente a manera de anécdota\cite{Martin}, que este problema fue planteado con algunos valores concretos al matemático húngaro Von Neumann quien lo resolvió al ins\-tan\-te, al preguntarle como lo había realizado dijo algo así como: sumando las distancias parciales, ¡no conozco otro método! Lo sorprendente de la res\-pu\-es\-ta  otorgada por Neumann es que se esperaba dijera que lo había hecho aplicando la ecuación (\ref{dis}) y no realizando una suma infinita de distancias parciales \cite{Martin}. 

En esta sección se analiza el problema mediante el enfoque de Neumann, es decir orientando la atención en las distancias parciales que recorre la mosca al pasearse entre los trenes, pues sumando estas distancias se tendrá la distancia total recorrida por la mosca. Para ello, es preciso hacer infinitas sumas ya que al considerar la mosca como un punto matemático y al despreciar el tiempo que tarda en girar al tocar un tren y dirigirse al otro cada vez tiene que recorrer menos distancia con igual rapidez, lo que  trae como consecuencia que el número de viajes tienda  a infinito mientras la distancia entre los trenes tiende a cero.  

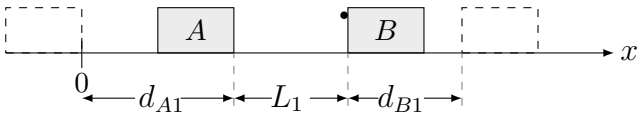
\begin{figure}[ht]
\begin{center}
\begin{tikzpicture}[xscale=1.0,yscale=1.0]
\draw[dashed,color=black] (0,0) rectangle (1,0.6);
\fill[gray!15](2,0) rectangle (3,0.6);
\draw[color=black] (2,0) rectangle (3,0.6);
\node[] at (2.5,0.3) {\small{$A$}};
\draw[dashed, color=black] (6,0) rectangle (7,0.6);
\fill[gray!15] (4.5,0) rectangle (5.5,0.6);
\draw[color=black] (4.5,0) rectangle (5.5,0.6);
\node[] at (5.0,0.3) {\small{$B$}};
\draw[-latex, black] (0cm,0cm)--(8cm,0cm);
\draw[black] (1cm,-0.2cm)--(1cm,0.2cm);
\node[] at (1.0,-0.4) {\small{$0$}};
\node[] at (8.2,0.0) {$x$};
\fill[black] (4.45cm,0.5cm) circle(0.05cm); 
\draw[-latex, black] (1.7cm,-0.6cm)--(1.0cm,-0.6cm);
\draw[-latex, black] (2.3cm,-0.6cm)--(3.0cm,-0.6cm);
\node[] at (2.05,-0.6) {$d_{A1}$};
\draw[dashed, gray] (3cm,-0.7cm)--(3cm,0.0cm);
\draw[-latex, black] (3.5cm,-0.6cm)--(3.0cm,-0.6cm);
\draw[-latex, black] (4.0cm,-0.6cm)--(4.5cm,-0.6cm);
\node[] at (3.7,-0.6) {$L_{1}$};
\draw[dashed, gray] (4.5cm,-0.7cm)--(4.5cm,0.0cm);
\draw[-latex, black] (4.9cm,-0.6cm)--(4.5cm,-0.6cm);
\draw[-latex, black] (5.5cm,-0.6cm)--(6.0cm,-0.6cm);
\node[] at (5.2,-0.6) {$d_{B1}$};
\draw[dashed, gray] (6cm,-0.7cm)--(6cm,0.0cm);
\end{tikzpicture}
\caption{Posición de la mosca y los dos trenes justo cuando la mosca toca el tren $B$ por primera vez.}
\label{Mosca2}
\end{center}
\end{figure}

Cuando la mosca parte del tren $A$ invierte un tiempo $t_{cB}$ para tocar el tren $B$,  por un pro\-ce\-di\-mi\-en\-ton completamente análogo al desarrollo en la sección ($2$) este tiempo queda determinado por la ecuación
\begin{equation}\label{cA}
t_{cB}=\frac{L_{0}}{v_{m}+v_{B}}.
\end{equation}
Ahora, en el tiempo $t_{cB}$ la mosca recorre una distancia
\begin{equation}\label{dm1}
d_{m1}=\frac{v_{m}L_{0}}{v_{m}+v_{B}}.
\end{equation}
Las posiciones de la mosca y los dos trenes justo cuando la mosca toca
el tren $B$ se representa en el esquema de la Fig.\ref{Mosca2}.

\begin{figure}[ht]
\begin{center}
\begin{tikzpicture}[xscale=1.0,yscale=1.0]
\fill[gray!15](2,0) rectangle (3,0.6);
\draw[color=black] (2,0) rectangle (3,0.6);
\node[] at (2.5,0.3) {\small{$A$}};
\fill[gray!15] (4.5,0) rectangle (5.5,0.6);
\draw[color=black] (4.5,0) rectangle (5.5,0.6);
\node[] at (5.0,0.3) {\small{$B$}};
\draw[-latex, black] (1.5cm,0cm)--(6.5cm,0cm);
\node[] at (3.0,-0.2) {\small{$0$}};
\node[] at (6.7,0.0) {$x$};
\fill[black] (3.07cm,0.5cm) circle(0.05cm); 
\draw[dashed, gray] (3cm,-0.7cm)--(3cm,-0.4cm);
\draw[-latex, black] (3.5cm,-0.6cm)--(3.0cm,-0.6cm);
\draw[-latex, black] (4.0cm,-0.6cm)--(4.5cm,-0.6cm);
\node[] at (3.7,-0.6) {$L_{2}$};
\draw[dashed, gray] (4.5cm,-0.7cm)--(4.5cm,0.0cm);
\end{tikzpicture}
\caption{Configuración inicial después de que la mosca regresa al tren $A$ por primera vez.}
\label{Mosca4}
\end{center}
\end{figure}
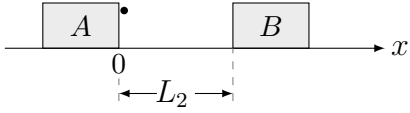

Para calcular las distancias recorridas por los dos trenes justo hasta el primer contacto se multiplica el tiempo obtenido en (\ref{cA}) por la rapidez de cada tren, así se obtienen las ecuaciones 
\begin{equation}\label{dA1}
d_{A1}=\frac{v_{A}L_{0}}{v_{m}+v_{B}}
\end{equation}
y
\begin{equation}\label{dB1}
d_{B1}=\frac{v_{B}L_{0}}{v_{m}+v_{B}}.
\end{equation}
Luego en el segundo recorrido, la mosca viaja desde el tren $B$ hasta el tren $A$ pero ahora la distancia de separación inicial entre los dos trenes está dada por
\begin{equation}\label{L1}
L_{1}=L_{0}-d_{A1}-d_{B1},
\end{equation}
sustituyendo (\ref{dA1}) y (\ref{dB1}) en (\ref{L1}) se obtiene
\begin{equation}\label{L12}
L_{1}=L_{0}\left(\frac{v_{m}-v_{A}}{v_{m}+v_{B}}\right).
\end{equation}
Continuando con el análisis, nótese que de acuerdo con la homogeneidad del espacio y el tiempo es posible definir un nuevo  origen de coordenadas y de tiempos \cite{Landau2}. Así, se reasigna la posición del tren $A$ como el origen en tiempo cero y $L_{1}$ pasa a ser la nueva posición inicial del tren $B$, ver Fig.\ref{Mosca3}. De esta forma el nuevo tiempo de contacto $t_{cA}$ entre la mosca y el tren $A$ viene dado por 
\begin{equation}\label{c2}
t_{cA}=\frac{L_{1}}{v_{m}+v_{A}}.
\end{equation}
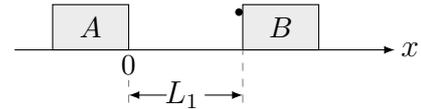
\begin{figure}[ht]
\begin{center}
\begin{tikzpicture}[xscale=1.0,yscale=1.0]
\fill[gray!15](2,0) rectangle (3,0.6);
\draw[color=black] (2,0) rectangle (3,0.6);
\node[] at (2.5,0.3) {\small{$A$}};
\fill[gray!15] (4.5,0) rectangle (5.5,0.6);
\draw[color=black] (4.5,0) rectangle (5.5,0.6);
\node[] at (5.0,0.3) {\small{$B$}};
\draw[-latex, black] (1.5cm,0cm)--(6.5cm,0cm);
\node[] at (3.0,-0.2) {\small{$0$}};
\node[] at (6.7,0.0) {$x$};
\fill[black] (4.45cm,0.5cm) circle(0.05cm); 
\draw[dashed, gray] (3cm,-0.7cm)--(3cm,-0.4cm);
\draw[-latex, black] (3.5cm,-0.6cm)--(3.0cm,-0.6cm);
\draw[-latex, black] (4.0cm,-0.6cm)--(4.5cm,-0.6cm);
\node[] at (3.7,-0.6) {$L_{1}$};
\draw[dashed, gray] (4.5cm,-0.7cm)--(4.5cm,0.0cm);
\end{tikzpicture}
\caption{Nueva configuración inicial.}
\label{Mosca3}
\end{center}
\end{figure}
Las ecuaciones que definen las distancias recorridas por la mosca y los trenes hasta el ins\-tan\-te en que
ocurre el segundo contacto, el decir cuando la mosca toca el tren $A$, son
\begin{equation}\label{m2}
d_{m2}=\frac{v_{m}L_{1}}{v_{m}+v_{A}},
\end{equation}
\begin{equation}\label{dA2}
d_{A2}=\frac{v_{A}L_{1}}{v_{m}+v_{A}}
\end{equation}
y
\begin{equation}\label{dB2}
d_{B2}=\frac{v_{B}L_{1}}{v_{m}+v_{A}}.
\end{equation}
Las nuevas posiciones de los trenes y la mosca así como la relación entre las distancias se muestra en la Fig.(\ref{Mosca4}).
Así, la nueva separación de los trenes está dada por
\begin{equation}\label{Ldos}
L_{2}=L_{1}-d_{A2}-d_{B2}.
\end{equation}
A partir de (\ref{c2}) y (\ref{Ldos}) se halla 
\begin{equation}\label{L2}
L_{2}=L_{1}\left(\frac{v_{m}-v_{B}}{v_{m}+v_{A}}\right).
\end{equation}
De forma iterativa se realiza este proceso, así que las ecuaciones (\ref{L1}) y (\ref{L2}) se renuevan en cada contacto de la mosca con  los trenes, para obtener la sucesión infinita 
\begin{multline}\label{suce}
\left\{L\right\}= \\
\left\{ L_{0},\alpha L_{0},\alpha \beta L_{0}, 
\alpha^{2}\beta L_{0},\alpha^{2}\beta^{2}L_{0}\ldots \right\},
\end{multline}
donde se ha definido $\alpha$ y $\beta$ como
\begin{equation}\label{ab}
\alpha=\frac{v_{m}-v_{A}}{v_{m}+v_{B}}\ ,\ \beta=\frac{v_{m}-v_{B}}{v_{m}+v_{A}}.
\end{equation}
Por otro lado, la suma $D$ de las distancias parciales es la distancia total recorrida por la
mosca, es decir 
\begin{equation}\label{e17}
D=\sum_{i=1}^{\infty}d_{mi}.
\end{equation}
De acuerdo con esto se puede escribir la distancia recorrida por la mosca separando los términos pares e impares de (\ref{e17}). Procediendo así y con ayuda de (\ref{suce}) se encuentra
\begin{multline}\label{pim}
D=B\left(\alpha_{0} \beta_{0} L_{0}+\beta\alpha L_{0}+\alpha^{2}\beta^{2}L_{0}+...\right)+\\
A\left(\alpha \beta^{0} L_{0}+\alpha^{2}\beta L_{0}+\alpha^{3}\beta^{2}L_{0}+...\right)
\end{multline} 
donde $A$ y $B$ estan definidos como
\begin{equation}\label{AB}
A=\frac{v_{m}}{v_{m}+v_{A}}\ ,\  B=\frac{v_{m}}{v_{m}+v_{B}}.
\end{equation}
Al expresar (\ref{pim}) abreviadamente se tiene
\begin{equation}\label{Abre}
D=L_{0}\left(B\sum_{k=0}^{\infty}\alpha^{k}\beta^{k}+A\sum_{k=0}^{\infty}\alpha^{k+1}\beta^{k}\right).
\end{equation}
Al extraer $\alpha$ del segundo término de la suma en (\ref{Abre}) se encuentra
\begin{equation}\label{suce2}
D=L_{0}\left(B\sum_{k=0}^{\infty}\alpha^{k}\beta^{k}+A \alpha \sum_{k=1}^{\infty}\alpha^{k}\beta^{k}\right).
\end{equation}
La expresión (\ref{suce2}) es una serie de geométrica  y por tanto converge dado que $\left|\alpha\beta\right|<1$, lo cual se espera ya que la rapidez de la mosca es menor que la rapidez de cada uno de los trenes. De esta manera (\ref{suce2}) se convierte en
\begin{equation}\label{DD}
D=L_{0}\left( \frac{B+\alpha A}{1-\alpha \beta}  \right).
\end{equation}
Al combinar las ecuaciones (\ref{ab}) y (\ref{AB})  se encuentra la relación 
\begin{equation}\label{alA}
\alpha A=B\left(\frac{v_{m}-v_{A}}{v_{m}+v_{A}} \right).
\end{equation}
Al sustituir \ref{alA} en \ref{DD}, la ecuación \ref{DD} toma la forma
\begin{equation}\label{DD2}
D=\frac{2L_{o}AB}{1-\alpha \beta}
\end{equation}
y observando que
\begin{equation}\label{DD3}
1-\alpha \beta=\frac{2AB\left(v_{A}+v_{B} \right)}{v_{m}}
\end{equation}
se obtiene
\begin{equation}\label{Dis}
D=\frac{{v_{m}L_{0}}}{v_{A}+v_{B}},
\end{equation}
ecuación que coincide con (\ref{dis}) obtenida manera e\-le\-men\-tal.

\section{Experimentos virtuales}

Para implementar una simulación que permita realizar un ExV sobre este sistema se asigna a cada tren una función de posición contra tiempo $x_A(t)$ y $x_B(t)$ así como de velocidad $v_A(t)$ y $v_B(t)$, donde el tiempo de evaluación va desde $t=0$ hasta $t=T_{max}$ incrementándose de $\Delta t$ en $\Delta t$, aquí $\Delta t$ es un tiempo pequeño comparado con el tiempo $T_{max}$ que se espera tarden los trenes en colisionar.  Por su parte la mosca tiene siempre asignada una rapidez constante $v_m$  y su velocidad cambia únicamente de signo cada vez que encuentra un tren. Así mismo,  el movimiento de los trenes y la mosca está regido por el algoritmo de Euler.

Este algoritmo  consiste en actualizar la posición de la mosca cada  paso temporal en $v_{m} \Delta t$, de ma\-ne\-ra que  la posición de la mosca se actualiza por cada incremento $\Delta t$ en
\begin{equation}\label{Euler}
 x_m \leftarrow \left(x_{m}+v_{m} \Delta t\right). 
\end{equation}
Dada la idealización del problema cada vez que la mosca toca  uno de los trenes cambia de manera instantánea su velocidad, esto se implementa  al cambiar el signo de su rapidez. Así
\begin{equation}\label{Euler2}
 -v_{m}\leftarrow v_{m} 
\end{equation}
a menos que la rapidez de la mosca sea menor que la rapidez del tren en cuyo caso sería capturada por este.  
\subsection{Algoritmo}
El algoritmo que materiza el ExV es el siguiente:
\begin{enumerate}
\item Se asignan los parámetros $T_{max}$, $\Delta t$, entre otros, y junto con estos las funciones $x_A(t)$, $x_B(t)$, $v_A(t)$ y $v_B(t)$
\item Se asignan las condiciones iniciales $x_{mo}$, $x_{Ao}$, $x_{Bo}$, $v_m=v_{mo}$, $v_{Ao}$ y $v_{Bo}$.
\item ¿Es $t\leq T_{max}$? Si la res\-pu\-es\-ta es afirmativa se va al paso $4$ de lo contrario se va al paso $10$.
\item Imprime $t$, $x_A$, $x_B$ y $x_m$.
\item Evalua la posición de la mosca mediante (\ref{Euler}) y evalua la posición y la velocidad de los trenes mediante $x_A(t)$, $x_B(t)$,$v_A(t)$ y $v_B(t)$.

\item Hace la pregunta $1$:¿es $x_m\geq x_B$? si la res\-pu\-es\-ta es afirmativa realiza la pregunta $2$: ¿es $v_m < v_B$? si la res\-pu\-es\-ta a la pregunta $2$ es afirmativa hace $v_{m}=v_{B}$ y va al paso $7$. Si la res\-pu\-es\-ta a la pregunta $2$ es negativa ejecuta (\ref{Euler2}), hace un paso mediante (\ref{Euler}) y va al paso $7$. Si la res\-pu\-es\-ta a la pregunta $1$ es negativa va directamente al paso $7$.  

\item Hace la pregunta $1$:¿es $x_m \leq x_A$? si la res\-pu\-es\-ta es afirmativa realiza la pregunta $2$: ¿es $v_m < v_A$? si la res\-pu\-es\-ta a la pregunta $2$ es afirmativa hace $v_{m}=v_{A}$ y va al paso $8$. Si la res\-pu\-es\-ta a la pregunta $2$ es negativa ejecuta (\ref{Euler2}), hace un paso mediante (\ref{Euler}) y va al paso $8$. Si la res\-pu\-es\-ta a la pregunta $1$ es negativa va directamente al paso $8$.

\item ¿Es $x_A \geq x_B$? si la res\-pu\-es\-ta es afirmativa va al paso $10$. Si la res\-pu\-es\-ta es negativa va  al paso $9$. 
\item Asigna $t\leftarrow \left(t+\Delta t\right)$ y retorna al paso $3$.
\item Finaliza.
\end{enumerate}

\subsection{ExV sobre el problema tradicional}
Se muestra un caso particular del problema tradicional el cual se rige  con los parámetros mostrados en la tabla (\ref{tab1}) y las condiciones iniciales mostradas en la tabla (\ref{tab2}).
\begin{table}[htp]
\caption{Parámetros.}\label{tab1} 
\vspace{2mm}
\centering
\begin{tabular}{|c|c|}
\hline  
 $T_{max}$ & $\Delta t$    \\ \hline  \hline
 
 $10s$   & $0,001s$   \\ \hline  
\end{tabular}
\end{table}
\begin{table}[htp]
\caption{Condiciones iniciales.}\label{tab2} 
\vspace{2mm}
\centering
\begin{tabular}{|c|c|c|c|c|c|}
\hline  
 $x_{mo}$  & $x_{Ao}$  & $x_{Bo}$  & $v_m$  & $v_A$  & $v_B$  \\ \hline  \hline
 
 $-5m$   & $-5m$   & $10m$   & $4\frac{m}{s}$   & $1\frac{m}{s}$ & $-0,5\frac{m}{s}$ \\ \hline  
\end{tabular}
\end{table}
\begin {figure}
\begin{center}
\input{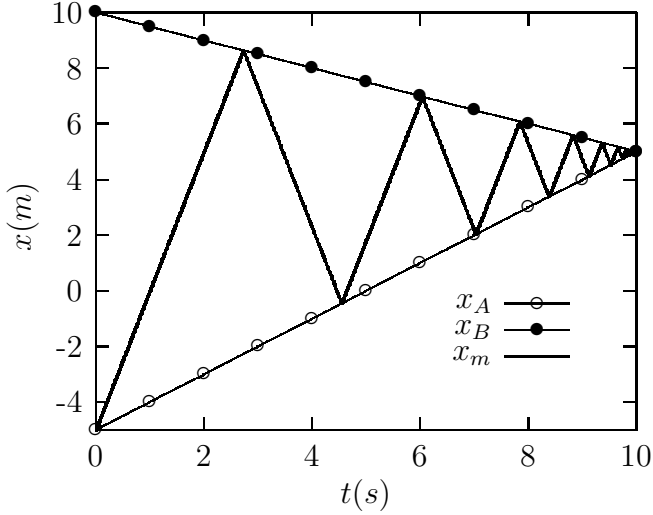}
\caption{ExV sobre el problema tradicional.}
\label{NeumannG}
\end{center}
\end {figure}

En la gráfica mostrada en la Fig.\ref{NeumannG} se observa  la mosca oscilando entre los trenes, aquí se puede leer el tiempo y la posición estimados para el  contacto de la mosca con cada tren, así mismo se puede estimar  el tiempo de colisión de los trenes y evidenciar algunos encuentros entre la mosca y los trenes. 

\subsection{ExV con trenes ace\-le\-ra\-dos}
Como un ejemplo de generalización del movimiento de los trenes considérese una situación en la cual la posición del tren $A$ está dada por 
\begin{equation}\label{ArA}
 x_{A}=x_{Ao}+v_{Ao}t+\frac{a}{2}t^{2} 
\end{equation}
y la del tren $B$ por
\begin{equation}\label{ArB}
 x_{B}=x_{Bo}e^{-\gamma t}. 
\end{equation}
Además, la mosca se sigue moviendo entre los trenes con rapidez constante $v_m$ y parte de la poción $x_{mo}=0m$. Los parámetros y condiciones iníciales del sistema se muestran en las tablas \ref{tab3} y \ref{tab4}, respectivamente. 
\begin{table}[h]
\caption{Parámetros para trenes ace\-le\-ra\-dos.}\label{tab3} 
\vspace{2mm}
\centering
\begin{tabular}{|c|c|c|c|}
\hline  
 $T_{max}$ & $\Delta t$ & $\gamma$  &  $a$  \\ \hline  \hline
 
 $14s$   & $0,01s$  & $0,4s^{-1}$  & $0,5\frac{m}{s^2}$ \\ \hline  
\end{tabular}
\end{table}
\begin{table}[h]
\caption{Condiciones iniciales para trenes ace\-le\-ra\-dos.}
\label{tab4} 
\vspace{2mm}
\centering
\begin{tabular}{|c|c|c|c|c|c|}
\hline  
 $x_{mo}$  & $x_{Ao}$  & $x_{Bo}$  & $v_m$  & $v_{Ao}$  & $v_{Bo}$  \\ \hline  \hline
 
 $0m$      & $-5m$     & $10m$     & $5\frac{m}{s}$   & $-3\frac{m}{s}$ & $-2\frac{m}{s}$ \\ \hline  
\end{tabular}
\end{table}
\begin {figure}
\begin{center}
\input{Fly.tex}
\caption{ExV sobre el problema con trenes ace\-le\-ra\-dos.}
\label{NeumannAr}
\end{center}
\end {figure}

En la Fig.\ref{NeumannAr} se muestran los resultados del ExV implementado para visualizar el movimiento de la mosca entre los trenes que se mueven de acuerdo con las ecuaciones (\ref{ArA}) y (\ref{ArB}).

\section{La mosca como  sensor de velocidad}
A ma\-ne\-ra de sugerencia didáctica se muestra como este  problema ideal  tiene una analogía física asociada al campo de las aplicaciones técnicas, lo cual se entiende al  realizar una correspondencia entre el problema de la mosca y un pulso de ultrasonido que se mueve entre los trenes. 

Ignorando los detalles técnicos de  emisión y recepción así como las correcciones por efecto Do\-ppler  de los pulsos de ultrasonido  el ExV  permite ilustrar el principio físico de algunos sensores de velocidad, para esto se con\-si\-de\-ra el tren $A$ en reposo y la mosca se identifica con el pulso, es decir $x_{m}\leftarrow x_{p}$. Tal pulso viaja hasta encontrar el tren $B$, al llegar allí se refleja y retorna al tren $A$, donde este emite simultáneamente con la llegada del pulso otro pulso en dirección del tren $B$ que a su vez se refleja y regresa de nuevo al tren $A$, repitiendose este proceso  muchas veces. 

Como ejemplo considere el tren A en reposo y al el tren B moviéndose con rapidez constante hacia el tren A. Al aplicar el ExV se encuentra un resultado como el mostrado en la Fig.\ref{Cens}. Si ahora se procesan los  resultados virtuales  capturados de posición contra tiempo y se organizan en una gráfica de posición contra tiempo se obtendrá una gráfica como la mostrada  en la Fig.\ref{Cens2}, donde se observa que la toma virtual de datos se hace cada vez en menor tiempo debido a que el tren $B$ está cada vez más cerca al tren $A$ y el pulso recorre menos distancia. Nótese que de Fig.\ref{Cens2} se puede inferir que la rapidez del tren $B$ es $\approx33,3m/s$.

El anterior pro\-ce\-di\-mi\-en\-ton se puede generalizar para estimar la rapidez del tren $B$ cualquiera sea su movimiento. 

\begin {figure}[h]
\begin{center}
\input{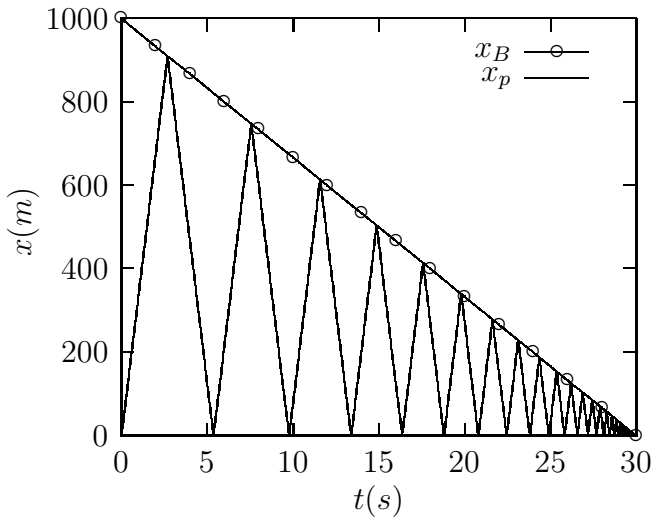}
\caption{Sesor de velocidad.}
\label{Cens}
\end{center}
\end {figure}
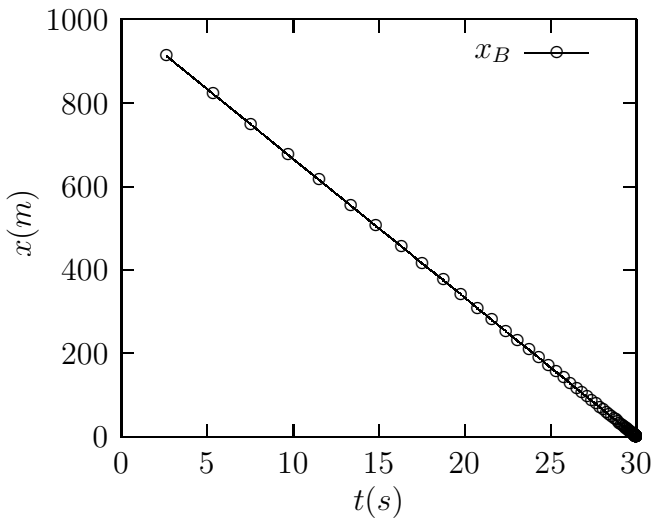
\begin {figure}[htp]
\begin{center}
\input{censor2.tex}
\caption{Datos de posición en el contacto.}
\label{Cens2}
\end{center}
\end {figure}

\section{Conclusiones}
Se repasó el problema tradicional de una mosca que se mueve entre dos trenes, describiendo tanto  una solución trivial como una  solución tipo Neumann, se mostró como implementar  un ExV que permitió generalizar este problema y aplicar dichos análisis a la explicación del principio físico de los sensores de velocidad basados en ultrasonido, brindando así un contexto de discusión sobre la interpretación gráfica en la cinemática unidimensional, el concepto de experimentó mental y la relación físico-matemática. Se propuso un algoritmo simple y fácil de implementar ya sea en una hoja de cálculo o en un lenguaje de programación, tal algoritmo permitió realizar experimentación virtual sobre este problema. Se espera que problemas similares de contenido introductorio sean abordados mediante esta técnica posibilitando el desarrollo de la intuición física y el acercamiento a la física computacional a aquellos  estudiantes que se inician en carreras de ciencias o ingeniería.    

\section*{Agradecimientos}

Los autores agradecen a la profesora Fernanda Santana y al profesor Guillermo Avendaño por la atenta lectura del manuscrito; a los estudiantes que han participado en la discusión del problema de la mosca en sus diversas presentaciones; al Departamento de Ciencias Naturales de la Universidad  Central por el apoyo y el tiempo asignado a la investigación y al CICATA del IPN de México por su continua colaboración.

\renewcommand{\refname}{Referencias}

\end{document}

%% file: censor2.tex
\setlength{\unitlength}{0.240900pt}
\ifx\plotpoint\undefined\newsavebox{\plotpoint}\fi
\sbox{\plotpoint}{\rule[-0.200pt]{0.400pt}{0.400pt}}%
\begin{picture}(1062,826)(0,0)
\sbox{\plotpoint}{\rule[-0.200pt]{0.400pt}{0.400pt}}%
\put(211.0,131.0){\rule[-0.200pt]{4.818pt}{0.400pt}}
\put(191,131){\makebox(0,0)[r]{ 0}}
\put(991.0,131.0){\rule[-0.200pt]{4.818pt}{0.400pt}}
\put(211.0,262.0){\rule[-0.200pt]{4.818pt}{0.400pt}}
\put(191,262){\makebox(0,0)[r]{ 200}}
\put(991.0,262.0){\rule[-0.200pt]{4.818pt}{0.400pt}}
\put(211.0,393.0){\rule[-0.200pt]{4.818pt}{0.400pt}}
\put(191,393){\makebox(0,0)[r]{ 400}}
\put(991.0,393.0){\rule[-0.200pt]{4.818pt}{0.400pt}}
\put(211.0,523.0){\rule[-0.200pt]{4.818pt}{0.400pt}}
\put(191,523){\makebox(0,0)[r]{ 600}}
\put(991.0,523.0){\rule[-0.200pt]{4.818pt}{0.400pt}}
\put(211.0,654.0){\rule[-0.200pt]{4.818pt}{0.400pt}}
\put(191,654){\makebox(0,0)[r]{ 800}}
\put(991.0,654.0){\rule[-0.200pt]{4.818pt}{0.400pt}}
\put(211.0,785.0){\rule[-0.200pt]{4.818pt}{0.400pt}}
\put(191,785){\makebox(0,0)[r]{ 1000}}
\put(991.0,785.0){\rule[-0.200pt]{4.818pt}{0.400pt}}
\put(211.0,131.0){\rule[-0.200pt]{0.400pt}{4.818pt}}
\put(211,90){\makebox(0,0){$0$}}
\put(211.0,765.0){\rule[-0.200pt]{0.400pt}{4.818pt}}
\put(344.0,131.0){\rule[-0.200pt]{0.400pt}{4.818pt}}
\put(344,90){\makebox(0,0){$5$}}
\put(344.0,765.0){\rule[-0.200pt]{0.400pt}{4.818pt}}
\put(478.0,131.0){\rule[-0.200pt]{0.400pt}{4.818pt}}
\put(478,90){\makebox(0,0){$10$}}
\put(478.0,765.0){\rule[-0.200pt]{0.400pt}{4.818pt}}
\put(611.0,131.0){\rule[-0.200pt]{0.400pt}{4.818pt}}
\put(611,90){\makebox(0,0){$15$}}
\put(611.0,765.0){\rule[-0.200pt]{0.400pt}{4.818pt}}
\put(744.0,131.0){\rule[-0.200pt]{0.400pt}{4.818pt}}
\put(744,90){\makebox(0,0){$20$}}
\put(744.0,765.0){\rule[-0.200pt]{0.400pt}{4.818pt}}
\put(878.0,131.0){\rule[-0.200pt]{0.400pt}{4.818pt}}
\put(878,90){\makebox(0,0){$25$}}
\put(878.0,765.0){\rule[-0.200pt]{0.400pt}{4.818pt}}
\put(1011.0,131.0){\rule[-0.200pt]{0.400pt}{4.818pt}}
\put(1011,90){\makebox(0,0){$30$}}
\put(1011.0,765.0){\rule[-0.200pt]{0.400pt}{4.818pt}}
\put(211.0,131.0){\rule[-0.200pt]{0.400pt}{157.549pt}}
\put(211.0,131.0){\rule[-0.200pt]{192.720pt}{0.400pt}}
\put(1011.0,131.0){\rule[-0.200pt]{0.400pt}{157.549pt}}
\put(211.0,785.0){\rule[-0.200pt]{192.720pt}{0.400pt}}
\put(50,458){\makebox(0,0){$\rotatebox{90} { \parbox{4cm}{\begin{align*}  x(m)  \end{align*}}}$}}
\put(611,29){\makebox(0,0){$t(s)$}}
\put(818,732){\makebox(0,0)[r]{$x_{B}$}}
\put(838.0,732.0){\rule[-0.200pt]{24.090pt}{0.400pt}}
\put(282,727){\usebox{\plotpoint}}
\multiput(282.00,725.92)(0.610,-0.499){115}{\rule{0.588pt}{0.120pt}}
\multiput(282.00,726.17)(70.779,-59.000){2}{\rule{0.294pt}{0.400pt}}
\multiput(354.00,666.92)(0.615,-0.498){93}{\rule{0.592pt}{0.120pt}}
\multiput(354.00,667.17)(57.772,-48.000){2}{\rule{0.296pt}{0.400pt}}
\multiput(413.00,618.92)(0.604,-0.498){93}{\rule{0.583pt}{0.120pt}}
\multiput(413.00,619.17)(56.789,-48.000){2}{\rule{0.292pt}{0.400pt}}
\multiput(471.00,570.92)(0.615,-0.498){75}{\rule{0.592pt}{0.120pt}}
\multiput(471.00,571.17)(46.771,-39.000){2}{\rule{0.296pt}{0.400pt}}
\multiput(519.00,531.92)(0.613,-0.498){77}{\rule{0.590pt}{0.120pt}}
\multiput(519.00,532.17)(47.775,-40.000){2}{\rule{0.295pt}{0.400pt}}
\multiput(568.00,491.92)(0.609,-0.497){61}{\rule{0.588pt}{0.120pt}}
\multiput(568.00,492.17)(37.781,-32.000){2}{\rule{0.294pt}{0.400pt}}
\multiput(607.00,459.92)(0.625,-0.497){61}{\rule{0.600pt}{0.120pt}}
\multiput(607.00,460.17)(38.755,-32.000){2}{\rule{0.300pt}{0.400pt}}
\multiput(647.00,427.92)(0.592,-0.497){51}{\rule{0.574pt}{0.120pt}}
\multiput(647.00,428.17)(30.808,-27.000){2}{\rule{0.287pt}{0.400pt}}
\multiput(679.00,400.92)(0.635,-0.497){49}{\rule{0.608pt}{0.120pt}}
\multiput(679.00,401.17)(31.739,-26.000){2}{\rule{0.304pt}{0.400pt}}
\multiput(712.00,374.92)(0.614,-0.496){41}{\rule{0.591pt}{0.120pt}}
\multiput(712.00,375.17)(25.774,-22.000){2}{\rule{0.295pt}{0.400pt}}
\multiput(739.00,352.92)(0.591,-0.496){41}{\rule{0.573pt}{0.120pt}}
\multiput(739.00,353.17)(24.811,-22.000){2}{\rule{0.286pt}{0.400pt}}
\multiput(765.00,330.92)(0.611,-0.495){33}{\rule{0.589pt}{0.119pt}}
\multiput(765.00,331.17)(20.778,-18.000){2}{\rule{0.294pt}{0.400pt}}
\multiput(787.00,312.92)(0.611,-0.495){33}{\rule{0.589pt}{0.119pt}}
\multiput(787.00,313.17)(20.778,-18.000){2}{\rule{0.294pt}{0.400pt}}
\multiput(809.00,294.92)(0.600,-0.494){27}{\rule{0.580pt}{0.119pt}}
\multiput(809.00,295.17)(16.796,-15.000){2}{\rule{0.290pt}{0.400pt}}
\multiput(827.00,279.92)(0.644,-0.494){25}{\rule{0.614pt}{0.119pt}}
\multiput(827.00,280.17)(16.725,-14.000){2}{\rule{0.307pt}{0.400pt}}
\multiput(845.00,265.92)(0.625,-0.492){21}{\rule{0.600pt}{0.119pt}}
\multiput(845.00,266.17)(13.755,-12.000){2}{\rule{0.300pt}{0.400pt}}
\multiput(860.00,253.92)(0.576,-0.493){23}{\rule{0.562pt}{0.119pt}}
\multiput(860.00,254.17)(13.834,-13.000){2}{\rule{0.281pt}{0.400pt}}
\multiput(875.00,240.92)(0.600,-0.491){17}{\rule{0.580pt}{0.118pt}}
\multiput(875.00,241.17)(10.796,-10.000){2}{\rule{0.290pt}{0.400pt}}
\multiput(887.00,230.93)(0.669,-0.489){15}{\rule{0.633pt}{0.118pt}}
\multiput(887.00,231.17)(10.685,-9.000){2}{\rule{0.317pt}{0.400pt}}
\multiput(899.00,221.93)(0.553,-0.489){15}{\rule{0.544pt}{0.118pt}}
\multiput(899.00,222.17)(8.870,-9.000){2}{\rule{0.272pt}{0.400pt}}
\multiput(909.00,212.93)(0.626,-0.488){13}{\rule{0.600pt}{0.117pt}}
\multiput(909.00,213.17)(8.755,-8.000){2}{\rule{0.300pt}{0.400pt}}
\multiput(919.00,204.93)(0.569,-0.485){11}{\rule{0.557pt}{0.117pt}}
\multiput(919.00,205.17)(6.844,-7.000){2}{\rule{0.279pt}{0.400pt}}
\multiput(927.00,197.93)(0.671,-0.482){9}{\rule{0.633pt}{0.116pt}}
\multiput(927.00,198.17)(6.685,-6.000){2}{\rule{0.317pt}{0.400pt}}
\multiput(935.00,191.93)(0.581,-0.482){9}{\rule{0.567pt}{0.116pt}}
\multiput(935.00,192.17)(5.824,-6.000){2}{\rule{0.283pt}{0.400pt}}
\multiput(942.00,185.93)(0.710,-0.477){7}{\rule{0.660pt}{0.115pt}}
\multiput(942.00,186.17)(5.630,-5.000){2}{\rule{0.330pt}{0.400pt}}
\multiput(949.00,180.93)(0.599,-0.477){7}{\rule{0.580pt}{0.115pt}}
\multiput(949.00,181.17)(4.796,-5.000){2}{\rule{0.290pt}{0.400pt}}
\multiput(955.00,175.94)(0.627,-0.468){5}{\rule{0.600pt}{0.113pt}}
\multiput(955.00,176.17)(3.755,-4.000){2}{\rule{0.300pt}{0.400pt}}
\multiput(960.00,171.94)(0.627,-0.468){5}{\rule{0.600pt}{0.113pt}}
\multiput(960.00,172.17)(3.755,-4.000){2}{\rule{0.300pt}{0.400pt}}
\multiput(965.00,167.94)(0.481,-0.468){5}{\rule{0.500pt}{0.113pt}}
\multiput(965.00,168.17)(2.962,-4.000){2}{\rule{0.250pt}{0.400pt}}
\multiput(969.00,163.95)(0.685,-0.447){3}{\rule{0.633pt}{0.108pt}}
\multiput(969.00,164.17)(2.685,-3.000){2}{\rule{0.317pt}{0.400pt}}
\multiput(973.00,160.95)(0.685,-0.447){3}{\rule{0.633pt}{0.108pt}}
\multiput(973.00,161.17)(2.685,-3.000){2}{\rule{0.317pt}{0.400pt}}
\put(977,157.17){\rule{0.700pt}{0.400pt}}
\multiput(977.00,158.17)(1.547,-2.000){2}{\rule{0.350pt}{0.400pt}}
\multiput(980.00,155.95)(0.462,-0.447){3}{\rule{0.500pt}{0.108pt}}
\multiput(980.00,156.17)(1.962,-3.000){2}{\rule{0.250pt}{0.400pt}}
\put(983,152.17){\rule{0.482pt}{0.400pt}}
\multiput(983.00,153.17)(1.000,-2.000){2}{\rule{0.241pt}{0.400pt}}
\put(985,150.17){\rule{0.700pt}{0.400pt}}
\multiput(985.00,151.17)(1.547,-2.000){2}{\rule{0.350pt}{0.400pt}}
\put(988,148.17){\rule{0.482pt}{0.400pt}}
\multiput(988.00,149.17)(1.000,-2.000){2}{\rule{0.241pt}{0.400pt}}
\put(990,146.67){\rule{0.482pt}{0.400pt}}
\multiput(990.00,147.17)(1.000,-1.000){2}{\rule{0.241pt}{0.400pt}}
\put(992,145.17){\rule{0.482pt}{0.400pt}}
\multiput(992.00,146.17)(1.000,-2.000){2}{\rule{0.241pt}{0.400pt}}
\put(994,143.67){\rule{0.241pt}{0.400pt}}
\multiput(994.00,144.17)(0.500,-1.000){2}{\rule{0.120pt}{0.400pt}}
\put(995,142.67){\rule{0.482pt}{0.400pt}}
\multiput(995.00,143.17)(1.000,-1.000){2}{\rule{0.241pt}{0.400pt}}
\put(997,141.67){\rule{0.241pt}{0.400pt}}
\multiput(997.00,142.17)(0.500,-1.000){2}{\rule{0.120pt}{0.400pt}}
\put(998,140.67){\rule{0.241pt}{0.400pt}}
\multiput(998.00,141.17)(0.500,-1.000){2}{\rule{0.120pt}{0.400pt}}
\put(999,139.67){\rule{0.241pt}{0.400pt}}
\multiput(999.00,140.17)(0.500,-1.000){2}{\rule{0.120pt}{0.400pt}}
\put(1000,138.67){\rule{0.241pt}{0.400pt}}
\multiput(1000.00,139.17)(0.500,-1.000){2}{\rule{0.120pt}{0.400pt}}
\put(1001,137.67){\rule{0.241pt}{0.400pt}}
\multiput(1001.00,138.17)(0.500,-1.000){2}{\rule{0.120pt}{0.400pt}}
\put(1002,136.67){\rule{0.241pt}{0.400pt}}
\multiput(1002.00,137.17)(0.500,-1.000){2}{\rule{0.120pt}{0.400pt}}
\put(1004,135.67){\rule{0.241pt}{0.400pt}}
\multiput(1004.00,136.17)(0.500,-1.000){2}{\rule{0.120pt}{0.400pt}}
\put(1003.0,137.0){\usebox{\plotpoint}}
\put(1005,136){\usebox{\plotpoint}}
\put(1005,134.67){\rule{0.241pt}{0.400pt}}
\multiput(1005.00,135.17)(0.500,-1.000){2}{\rule{0.120pt}{0.400pt}}
\put(1006,135){\usebox{\plotpoint}}
\put(1006.0,135.0){\usebox{\plotpoint}}
\put(1007.0,134.0){\usebox{\plotpoint}}
\put(1007.0,134.0){\usebox{\plotpoint}}
\put(1008.0,133.0){\usebox{\plotpoint}}
\put(1008.0,133.0){\usebox{\plotpoint}}
\put(1009.0,132.0){\usebox{\plotpoint}}
\put(1009.0,132.0){\usebox{\plotpoint}}
\put(1010.0,131.0){\usebox{\plotpoint}}
\put(282,727){\makebox(0,0){$\circ$}}
\put(354,668){\makebox(0,0){$\circ$}}
\put(413,620){\makebox(0,0){$\circ$}}
\put(471,572){\makebox(0,0){$\circ$}}
\put(519,533){\makebox(0,0){$\circ$}}
\put(568,493){\makebox(0,0){$\circ$}}
\put(607,461){\makebox(0,0){$\circ$}}
\put(647,429){\makebox(0,0){$\circ$}}
\put(679,402){\makebox(0,0){$\circ$}}
\put(712,376){\makebox(0,0){$\circ$}}
\put(739,354){\makebox(0,0){$\circ$}}
\put(765,332){\makebox(0,0){$\circ$}}
\put(787,314){\makebox(0,0){$\circ$}}
\put(809,296){\makebox(0,0){$\circ$}}
\put(827,281){\makebox(0,0){$\circ$}}
\put(845,267){\makebox(0,0){$\circ$}}
\put(860,255){\makebox(0,0){$\circ$}}
\put(875,242){\makebox(0,0){$\circ$}}
\put(887,232){\makebox(0,0){$\circ$}}
\put(899,223){\makebox(0,0){$\circ$}}
\put(909,214){\makebox(0,0){$\circ$}}
\put(919,206){\makebox(0,0){$\circ$}}
\put(927,199){\makebox(0,0){$\circ$}}
\put(935,193){\makebox(0,0){$\circ$}}
\put(942,187){\makebox(0,0){$\circ$}}
\put(949,182){\makebox(0,0){$\circ$}}
\put(955,177){\makebox(0,0){$\circ$}}
\put(960,173){\makebox(0,0){$\circ$}}
\put(965,169){\makebox(0,0){$\circ$}}
\put(969,165){\makebox(0,0){$\circ$}}
\put(973,162){\makebox(0,0){$\circ$}}
\put(977,159){\makebox(0,0){$\circ$}}
\put(980,157){\makebox(0,0){$\circ$}}
\put(983,154){\makebox(0,0){$\circ$}}
\put(985,152){\makebox(0,0){$\circ$}}
\put(988,150){\makebox(0,0){$\circ$}}
\put(990,148){\makebox(0,0){$\circ$}}
\put(992,147){\makebox(0,0){$\circ$}}
\put(994,145){\makebox(0,0){$\circ$}}
\put(995,144){\makebox(0,0){$\circ$}}
\put(997,143){\makebox(0,0){$\circ$}}
\put(998,142){\makebox(0,0){$\circ$}}
\put(999,141){\makebox(0,0){$\circ$}}
\put(1000,140){\makebox(0,0){$\circ$}}
\put(1001,139){\makebox(0,0){$\circ$}}
\put(1002,138){\makebox(0,0){$\circ$}}
\put(1003,137){\makebox(0,0){$\circ$}}
\put(1004,137){\makebox(0,0){$\circ$}}
\put(1005,136){\makebox(0,0){$\circ$}}
\put(1005,136){\makebox(0,0){$\circ$}}
\put(1006,135){\makebox(0,0){$\circ$}}
\put(1006,135){\makebox(0,0){$\circ$}}
\put(1007,135){\makebox(0,0){$\circ$}}
\put(1007,134){\makebox(0,0){$\circ$}}
\put(1007,134){\makebox(0,0){$\circ$}}
\put(1008,134){\makebox(0,0){$\circ$}}
\put(1008,133){\makebox(0,0){$\circ$}}
\put(1008,133){\makebox(0,0){$\circ$}}
\put(1009,133){\makebox(0,0){$\circ$}}
\put(1009,133){\makebox(0,0){$\circ$}}
\put(1009,133){\makebox(0,0){$\circ$}}
\put(1009,132){\makebox(0,0){$\circ$}}
\put(1009,132){\makebox(0,0){$\circ$}}
\put(1010,132){\makebox(0,0){$\circ$}}
\put(1010,132){\makebox(0,0){$\circ$}}
\put(1010,132){\makebox(0,0){$\circ$}}
\put(1010,132){\makebox(0,0){$\circ$}}
\put(1010,132){\makebox(0,0){$\circ$}}
\put(1010,132){\makebox(0,0){$\circ$}}
\put(1010,132){\makebox(0,0){$\circ$}}
\put(1010,132){\makebox(0,0){$\circ$}}
\put(1010,132){\makebox(0,0){$\circ$}}
\put(1010,131){\makebox(0,0){$\circ$}}
\put(1011,131){\makebox(0,0){$\circ$}}
\put(1011,131){\makebox(0,0){$\circ$}}
\put(1011,131){\makebox(0,0){$\circ$}}
\put(1011,131){\makebox(0,0){$\circ$}}
\put(1011,131){\makebox(0,0){$\circ$}}
\put(1011,131){\makebox(0,0){$\circ$}}
\put(1011,131){\makebox(0,0){$\circ$}}
\put(1011,131){\makebox(0,0){$\circ$}}
\put(1011,131){\makebox(0,0){$\circ$}}
\put(1011,131){\makebox(0,0){$\circ$}}
\put(1011,131){\makebox(0,0){$\circ$}}
\put(1011,131){\makebox(0,0){$\circ$}}
\put(1011,131){\makebox(0,0){$\circ$}}
\put(1011,131){\makebox(0,0){$\circ$}}
\put(1011,131){\makebox(0,0){$\circ$}}
\put(1011,131){\makebox(0,0){$\circ$}}
\put(1011,131){\makebox(0,0){$\circ$}}
\put(1011,131){\makebox(0,0){$\circ$}}
\put(1011,131){\makebox(0,0){$\circ$}}
\put(1011,131){\makebox(0,0){$\circ$}}
\put(1011,131){\makebox(0,0){$\circ$}}
\put(888,732){\makebox(0,0){$\circ$}}
\put(1010.0,131.0){\usebox{\plotpoint}}
\put(211.0,131.0){\rule[-0.200pt]{0.400pt}{157.549pt}}
\put(211.0,131.0){\rule[-0.200pt]{192.720pt}{0.400pt}}
\put(1011.0,131.0){\rule[-0.200pt]{0.400pt}{157.549pt}}
\put(211.0,785.0){\rule[-0.200pt]{192.720pt}{0.400pt}}
\end{picture}